# A Flexible Approach for Finding Optimal Paths with Minimal Conflicts*

J. Bowles and M. B. Caminati

School of Computer Science, University of St Andrews
KY16 9SX St Andrews, United Kingdom
{jkfb|mbc8}@st-andrews.ac.uk

**Abstract.** Complex systems are usually modelled through a combination of structural and behavioural models, where separate behavioural models make it easier to design and understand partial behaviour. When partial models are combined, we need to guarantee that they are consistent, and several automated techniques have been developed to check this. We argue that in some cases it is impossible to guarantee total consistency, and instead we want to find execution paths across such models with minimal conflicts with respect to a certain metric of interest. We present an efficient and scalable solution to find optimal paths through a combination of the theorem prover Isabelle with the constraint solver Z3. Our approach has been inspired by a healthcare problem, namely how to detect conflicts between medications taken by patients with multiple chronic conditions, and how to find preferable alternatives automatically.

## 1 Introduction

In complex systems design, it is common to model components separately in order to facilitate the understanding and analysis of their behaviour. Nonetheless, modelling the complete behaviour of a component is hard [24], and often sets of possible scenarios of execution are captured instead. A scenario describes a particular situation that may involve a component and how it behaves and interacts with other components. In practice, UML's sequence diagrams are commonly used to model scenarios [19]. From such individual scenarios, we then need to be able to derive the complete behaviour of a component. The same ideas apply if we model (partial) business processes within an organisation, for instance using BPMN [18]. In either case, we need a means to compose models (scenarios or processes), and when this cannot be done, detect and resolve inconsistencies.

Composing systems manually can only be done for small systems. As a result, in recent years, various methods for automated model composition have been introduced [1, 4, 21, 10, 12, 22, 25, 26, 27, 3, 5]. Most of these methods introduce algorithms to produce a composite model from partial specifications and assume a formal underlying semantics [10]. In our recent work [3, 5], we have used constraint solvers (Alloy [8] and Z3 [16] respectively) for automatically constructing the composed model. This involves generating all constraints associated to the models, and using an automated solver to find a solution (the composed model) for the conjunction of all constraints. Using Alloy for model composition (usually only for structural models), is an active area of research [22, 27], but the use of Z3 is a novelty of [5]. In [5] we did not exploit Z3's arithmetic capabilities which we have done more recently in [6]. Most existing approaches can detect inconsistencies, but fail to provide a means to resolve them.

---

* This research is supported by EPSRC grant EP/M014290/1.

We argue that in some cases it is impossible to guarantee total consistency between scenarios of execution. Instead we want to find execution paths across such models with minimal conflicts with respect to a certain metric of interest. We present an efficient and scalable solution to find optimal paths through a combination of the theorem prover Isabelle [17] with the constraint solver Z3. Our approach has been inspired by a healthcare problem, namely how to detect conflicts between medications taken by patients with multiple chronic conditions, and how to find preferable alternatives automatically. In this paper, we focus on the theoretical foundations required to address the problem and our medical domain, but remind the reader of the more general applicability of our work and considerable practical benefits.

This paper provides a formal statement of the problem, a measure of inconsistency, and shows how the obtained problem can be turned into an optimisation problem. In addition, it illustrates how a SMT solver can be used to find a scalable solution to the proposed problem. As a final contribution, the paper introduces a general technique to combine Z3 with Isabelle in order to ensure that the SMT translation of the problem is formally correct.

Paper structure: Section 2 introduces the formalisation of the problem and our solution. Section 3 translates this formulation into an SMT context. In Section 4, a concrete application in the healthcare domain is introduced to motivate the approach introduced in this paper, and is used to evaluate our design through a basic implementation featuring simple input and output interfaces. Section 5 exposes the general technique we used to combine the theorem prover Isabelle and the SMT solver to guarantee the correctness of the SMT code illustrated in Section 3. Section 6 discusses related work, and Section 7 concludes.

## 2 Description of the Problem and Approach

Our problem is formulated formally as follows: we are given a list of simple directed acyclic graphs $G_i, i = 1, \ldots, n$, each with exactly one source node. Since each of the graphs is simple, each $G_i$ can be thought of as a finite set of ordered pairs of nodes $(j, k)$, each representing a directed edge from node $j$ to node $k$. Therefore, we can define $G := \bigcup_{i=1}^{n} G_i$, and denote by $V(G')$ the set of nodes touched by any edge in $G' \subseteq G$. Further, we assume that $V(G_{i_1}) \cap V(G_{i_2}) = \emptyset$ for any $i_1 \neq i_2 \in \{i_1, \ldots, i_n\}$: i.e., distinct graphs have no nodes in common.

We want to obtain a list of paths, one for each given graph, each leading from the source of the corresponding graph to one of it sinks. Such a list of paths must be determined so as to maximise a given *score*, which can be thought of as a metric of the compatibility of the resulting execution paths. Before defining how this score is computed, we need to describe how the needed input data and the output are encoded. Examples later clarify the need for these notions.

## 2.1 Score model

To compute the score, we assume to be given further input, besides the list of graphs $G_1, \ldots, G_n$, as follows.

1. A map $t : G \to \mathbb{N} \times \mathbb{N}$ associating to each edge $e$ a pair $(t^-(e), t^+(e))$, where $t^-(e)$ is the minimal time and $t^+(e)$ is the maximal time $e$ has to wait before $e$ can occur. We require $t^-(e) \leq t^+(e)$ for any $e \in G$. It can be used to bound the occurrence of tasks associated to nodes in a graph. For example, $t^-$ expresses that the next task cannot start before $t^-$ time units, and $t^+$ expresses that it must occur before $t^+$ time units.
2. A list $\tau := \{\tau_1, \ldots, \tau_n\}$ of integers, where $\tau_i$ specifies the instant at which the source node of the graph $G_i$ is executed. This list can be used to express the requirement that different models start their execution at different times.
3. A finite set $R$ of *resources*.
4. A map $M : V(G) \to 2^R$. $M(j)$ specifies a subset of resources among which one can be chosen in order to perform the task corresponding to node $j$.
5. A map $g : R \to \mathbb{Z} \times \mathbb{N}$ associating to each resource $r$ an *effectiveness score* $g_1(r)$ and an *amount* $g_2(r)$. The effectiveness is a measure of how well a given resource performs a task, and the amount is how much of the resource is consumed for performing the task. For example, a hardware resource is needed for a given time, a medication must be taken at a given dosage, etc.
6. A map $I : R \times R \to \mathbb{Z}$ yielding an *interaction*. The interaction is an integer expressing how much two resources mutually boost or interfere, where a negative interaction means a counter-productive effect (i.e., diminishing the overall effectiveness of the two interacting resources).
7. A map $f : \mathbb{Z} \times \mathbb{N} \times \mathbb{N} \times \mathbb{N} \to \mathbb{Z}$ combining an interaction between two resources, a time distance, and the amounts of the two resources to yield the component of the overall score for a given pair of resources. $f$ takes into account the fact that the actual interaction between resources occurring at distinct nodes depends not only on the interaction between the resources (as defined at the previous point), but on their amount and on how much time passes between their occurrences. We will refer to the integer values returned by $f$ as *interaction scores*.

## 2.2 Output

Given a list $G_1, \ldots G_n$ of directed acyclic graphs, a set of resources $R$ and maps $f$, $g$, $I$, $M$, $t$ as introduced in the previous section, the output is a triple of functions $(F, c, m)$, each defined on the set of all nodes, $V(G)$.

$F$ is a boolean function telling us which nodes are executed, $c(j)$ returns the instant at which node $j$ is executed, while $m(j)$ is the resource picked to perform the task associated to the node $j$. We will use the notation $P[X]$ to indicate the image of the set $X$ through the relation $P$; for example, $F[\{\text{true}\}]$ is the set of executed nodes. With this notation in place, we require that $F$, $c$ and $m$ satisfy all the following conditions:

1. the set on which $F$ is true determines one path for each $G_i$, starting from the source of $G_i$ and ending at one of its sinks; this is a way to represent the paths that we anticipated as our main goal at the beginning of this section. More formally, $F$ satisfies the following requirement: for any $i \in \{1, \ldots, n\}$ there is a finite sequence $w^i$ of nodes of $G_i$ such that
   (a) $w_0^i$ is the source of $G_i$,
   (b) $w_{|w^i|-1}^i$ is a sink of $G_i$,
   (c) $\forall j \in \{1, \ldots, |w^i| - 1\}\ (w_{j-1}^i, w_j^i) \in G_i$, and
   (d) $\left\{w_0^i, \ldots, w_{|w^i|-1}^i\right\} = F^{-1}[\{\text{true}\}] \cap G_i$.
2. $\forall j \in V(G),\ F(j) \to m(j) \in M(j)$;
3. for any $i \in \{1, \ldots, n\}$, if $j$ is the source node of $G_i$, then $c(j) = \tau_i$;
4. if there is an edge going from the node $j$ to the node $k$, and $j$ and $k$ are executed, then
$$t^-(j,k) \leq c(k) - c(j) \leq t^+(j,k);$$
5. the global score:
$$\sum_{j \in F^{-1}[\{\text{true}\}]} g_1(m(j)) + \\ \sum_{\substack{i_1, i_2 \in \{1,\ldots,n\}, i_1 < i_2 \\ j \in F^{-1}[\{\text{true}\}] \cap V(G_{i_1}) \\ k \in F^{-1}[\{\text{true}\}] \cap V(G_{i_2})}} f(I(m(j), m(k)), |c(k) - c(j)|, g_2(m(j)), g_2(m(k))) \tag{1}$$

   is maximal.

The first term in expression (1) sums the effectiveness of each picked resource for each executed node, irrespectively of whether distinct picked resources interact. The second term sums together the interaction scores of each pair of resources picked in distinct graphs. The interaction score for each of such pairs depends on the absolute interaction of the two resources (specified by $I$), the time distance separating the occurrence of the two resources, and the amount.

### 2.3 An Illustrative Example

We use a simplified example to motivate our formal problem. A more realistic example and our solution is given in Section 4.

Assume that a patient with an acute condition is hospitalised on day 0. There are two possible treatments for the condition: a non-surgical treatment and surgery. The two alternatives are represented by the two branches in the directed graph of Fig. 1 (left), where the source node represents the hospitalisation. The right branch represents the choice of surgery with nodes $n_3$ and $n_4$ denoting the steps implied by this choice ($n_3$: pre-surgical testing and $n_4$: the surgery itself). Each node in a treatment graph may have one or more ways of performing it. For example, in the case of pre-surgical testing it involves administering one of two drugs ($d_1$ or $d_2$), while in the case of the surgery, only one resource is present (we assume here that there is only one way to perform it). The left branch (with $n_2$) models the non-surgical choice, here associated to the prescription of drug $d_0$ (with no other choice available). The weights on the edges are the time constraints for the subsequent step: for example, after pre-surgical testing was performed ($n_3$), surgery cannot happen before 2 days have passed, but should happen within 4 days (for illustration purposes only).

Additionally, this particular patient suffers from a chronic condition, which requires him to take drug $d_3$ on even days and $d_4$ on odd days. This can be represented by a path graph, unfolding the alternation of $d_3$ and $d_4$ for a given finite number of days (Fig. 1 right).

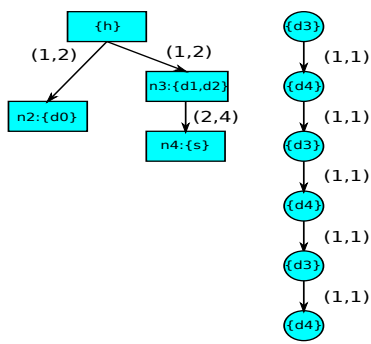

**Fig. 1.** A simple example problem.

It is known that surgery is preferred, but $d_1$ interacts negatively with $d_3$ and $d_2$ with $d_4$. We ignore drug dosages here, which our formalisation can handle as well. The problem we want to solve is to find how to best schedule steps in the hospitalisation, and how to choose between the non-surgical and the surgical treatment, taking into account the respective effectiveness and the interaction with the treatments for the chronic condition.

## 3 SMT Translation

We now need to represent the notions presented in Section 2 in a way amenable to SMT computations. To represent the output $F$, we introduce one boolean variable $\text{node}_i$ for each node $i$ (i.e., the truth value of $\text{node}_i$ will yield $F(i)$).

First, we must pick all the sources of the $G_i$'s, so that the corresponding node variables will be asserted to be true. Afterwards, for each source, we must assert that exactly one of its children must be true. Then, for every child set to true, we must ask that exactly one of its children must be true, and so on, until no node has children (we reached a sink). Besides doing that, we want to make sure that no other node is selected. Correspondingly, we generate, for the node $i$, the assertions

$$\text{node}_i \to \bigvee_{j \mid (i,j) \in G} \left( \left( \bigwedge_{\substack{k \neq j, \\ k \mid (i,k) \in G}} (\neg \text{node}_k) \right) \wedge \text{node}_j \right) \qquad (2)$$

$$\left( \bigwedge_{(j,i) \in G} (\neg \text{node}_j) \right) \to \neg \text{node}_i \qquad (3)$$

### 3.1 Scores in SMT

To represent the output $c$, we introduce one numerical variable $\text{clock}_j$ for each node $j$. For each $j, k$ such that $(j, k) \in G$, we create the following assertion:

$$\text{node}_j \wedge \text{node}_k \rightarrow \text{clock}_k - \text{clock}_j \geq t^-(j, k) \wedge \text{clock}_k - \text{clock}_j \leq t^+(j, k).$$

Additionally, we impose that each source node happens at the time specified by $\tau$; therefore, if $j$ is the source node of $G_i$, we assert:

$$\text{clock}_j = \tau_i. \qquad (4)$$

To represent the output $m$, we introduce one variable $\text{label}_j$ for each node $j$, whose value is the name of the drug picked for the node $j$. Now we can introduce two kinds of scores, represented by integer SMT variables $\text{score}_j$ and $\text{score}_{j,k}$, respectively: the first represents the effectiveness of the prescription associated to node $j$, while the second is the score generated by possible conflicts between the prescriptions picked for node $j$ and for node $k$. We first zero out the scores for the non-executed nodes:

$$\neg \text{node}_i \rightarrow \text{score}_{i,j} = 0 \wedge \text{score}_i = 0 \wedge \text{score}_{j,i} = 0.$$

Since the variables $\text{node}_j$ describe whether a node is executed the final optimisation, we generate the following assertions for each possible $j, k$:

$$\text{node}_j \wedge \text{node}_k \rightarrow \text{score}_{j,k} =$$
$$f\left(I\left(\text{label}_j, \text{label}_k\right), |\text{clock}_k - \text{clock}_j|, g_2\left(\text{label}_j\right), g_2\left(\text{label}_k\right)\right).$$

We finally assign the sum of the $\text{score}_{i,j}$ and of the $\text{score}_i$ to a variable, and ask for an SMT solution to all the assertions which maximises the global score. The map $I$ assigns a degree to the possible conflicts; this assignment, together with $f$, determine how the interactions influence the resulting choice of resources, timing, and the execution paths in the different models. These choices add flexibility to the whole approach but, on the other hand, need to be done by a domain expert.

## 4 Evaluation and Use Cases

In healthcare management and practice, as in other domains, clinical and medical procedures are streamlined by adopting standardised guidelines. In particular, treatments for common chronic conditions have been subject to various clinical trials, and the outcomes documented in *clinical pathways* specifying accepted treatment steps, possible alternatives, and recommendations to follow. Clinical pathways are informal flowcharts, with natural language annotations, and as such they can be formalised as directed acyclic graph structures, usually with one initial node (the source), and each node representing a medication prescription. Applying a single clinical pathway to a given patient is subject to a number of variable aspects and requirements, for example:

1. pathways typically present alternatives from which one is chosen;
2. there is often a choice to be taken among equivalent drugs in a group;
3. the time separating subsequent steps in a pathway is typically not fixed, being liable to be adapted to the context and the patient situation;
4. the dosage of the chosen drugs influences how drugs mutually interact;
5. the set of chronic conditions typically changes over the patient's life span: when this happens, even the treatment of the pre-existing conditions must be reconsidered.

On patients suffering from multiple chronic conditions, several pathways have to be applied concurrently, so that the number of possible combinations of these parameters increases dramatically. Our goal is to present the clinician suggestions about which choice of the parameters above is the best. The model presented in the previous sections allows us to capture all these aspects: the resources correspond to the single treatments (e.g., drugs, or surgical procedures), the effectiveness score expresses how well a single treatment performs, the amount can be used to express the dosage of a drug, and $I$ can model, for example, drug-drug interaction. For consistency and comparison, we evaluate our design on a well-known case [7] of a hypothetical 79-year-old woman with five diseases: chronic obstructive pulmonary disease (COPD), diabetes mellitus (type 2), hypertension, osteoarthritis, and osteoporosis. To this end, we extracted data to represent pathways and scores from two sources, respectively: NICE pathways[1], publicly available as informal flowcharts with accompanying text, and the website `drugs.com` to derive the scores.

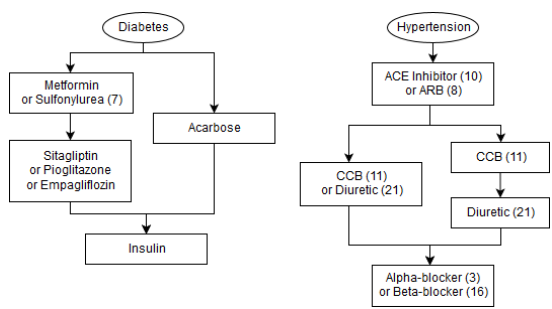

**Fig. 2.** Pharmaceutical graphs for Diabetes and Hypertension based on NICE pathways. ACE – angiotensin-converting-enzyme, ARB – angiotensin receptor blocker, CCB – calcium channel blocker.

Given the form in which pathways from NICE are presented, we extracted the data needed as input to our design manually, for each of the five conditions we are considering (see [11]). We pruned the nodes not liable to cause conflicts (e.g.,

---

[1] http://pathways.nice.org.uk/

"ongoing monitoring of $HbA_{1c}$"), generated an adjacency relation describing the underlying graph, and attached a list of possible medications for each node (for example, when a medication group such as *Sulfonylurea* was associated to a node, we inserted the list of all the medications in the group in the corresponding node of the graph we generated). The resulting graphs are visible in Fig. 2 and Fig. 3, where numbers in brackets represent the number of individual medications in a group. They contain a total of 127 distinct medications. Some of the graph parameters, such as the duration limits (described by the functions $t^-$ and $t^+$ introduced in Section 2) can depend on the single patient and on the context, and are therefore not present in the data provided by NICE. For evaluation purposes, we generated suitable data reasonable for our purposes to fill the gaps (this will be done by clinicians in the future). To retrieve all the possible drug conflicts, we used the interaction engine on the mentioned site `drugs.com`[2], and obtained a classification (minor, moderate and major): 178 minor conflicts, 3033 moderate conflicts, 270 major conflicts, for a total of 3481 conflicts, an amount too large to analyse manually.

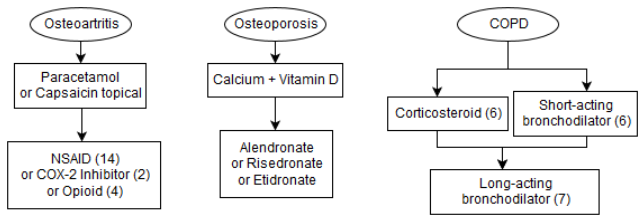

**Fig. 3.** Pharmaceutical graphs for three additional conditions. NSAID – nonsteroidal anti-inflammatory drug, COX-2 – Cyclooxygenase-2.

### 4.1 Results

The data extracted as explained above was represented in a textual, comma-separated value (CSV) format, and we built a simple implementation of our design, parsing these files, generating SMT code, producing text representations of $(F, c, m)$, together with a graphical representation of the output using Cytoscape [23]. For computations, we assigned the values of $-100$, $-1000$ and $-5000$ to minor, moderate and major conflicts, respectively. Another important parameter we had to set is $f$, which weights the conflicts according to time separation and dosage. We chose a simple form, which takes the conflict and zeroes it as soon as the time distance is greater than 8 time units or one of the dosages is less than 10 dosage units.

The output is shown in Fig. 4: the text in the nodes of Fig. 2 and Fig. 3 have been substituted by the drug ultimately picked in the maximised solution, except for the non-executed nodes, which are marked with "N/A". In the lower part of the interface, the user can see the node list with all the relevant information: the picked drug, the clock of the corresponding event, the score, the conflicting nodes (if any) and the conflict score (if any). The user can also click on single node to highlight only the relevant data. For this example, the total score is $-1220$, while the total conflict is $-2100$, composed as follows:

- the insulin-benazepril interaction contributes $-1000$,
- the doxazosin-fentanyl interaction contributes $-1000$,
- the benazepril-calciumVitaminD interaction contributes $-100$.

---
[2] http://www.drugs.com/drug_interactions.html

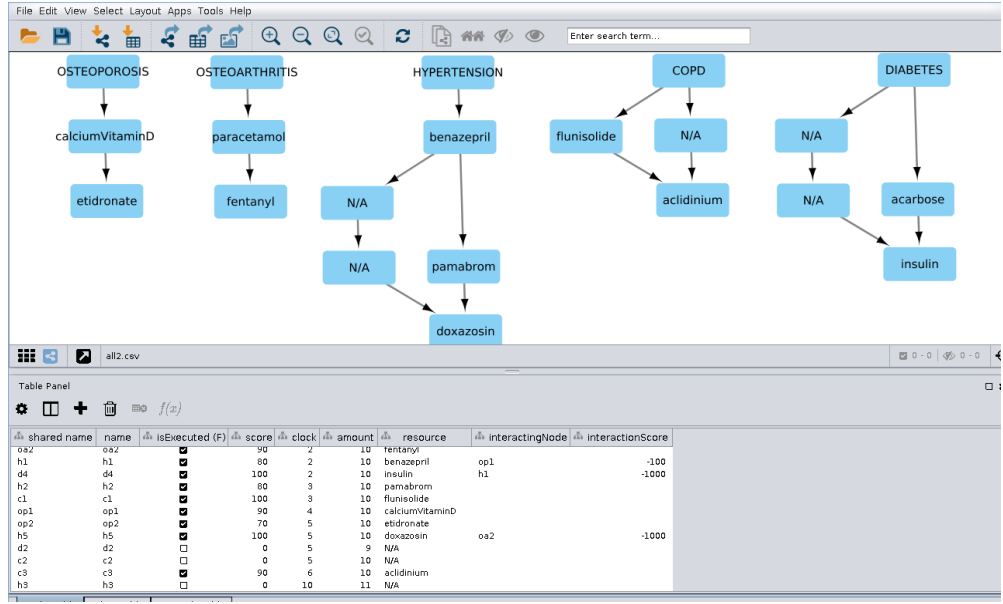

**Fig. 4.** Time- and dosage- aware optimal solution generated by Z3 for a hypothetical five-comorbid patient, as represented in the graphical front-end.

To assess the performance of the proposed approach, we timed the presented implementation. The average running time is 28.1 seconds, including the SMT code generation from the input data (which, however, takes a negligible amount of time); these results were obtained on an off-the-shelf laptop with 4GB RAM and a dual-core 2GHz CPU, running a 32-bit Linux OS. The actual run-times would likely be less, since in real situations a number of possibilities are excluded and additional restrictions often apply. For example, a portion of drugs could be excluded a priori because not available, or because known to have too many interactions; what is more, the doctor might impose manually a choice in a graph branch, thereby reducing the combinations.

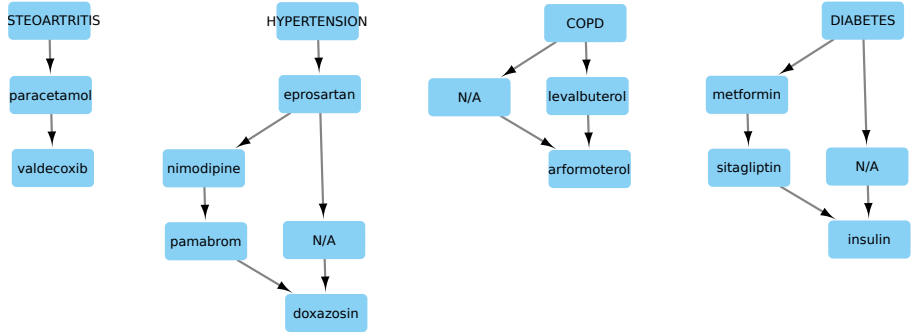

**Fig. 5.** Time- and dosage- aware optimal solution generated by Z3 with four concurrent morbidities.

### 4.2 Introducing time offsets

Suppose a patient is being treated for a number of conditions, when she gets diagnosed an additional condition (requirement (5) of Section 4). We want to show how the time-awareness of our design helps when facing such a situation. Let the instant in which the additional condition gets diagnosed be denoted by 0. Starting from time 0, we want to re-assess the patient's situation to get suggestion about which changes in the therapy should be implemented, assuming that the treatment for the pre-existing conditions started at some time in the past $-x$. To achieve that, it will suffice to change condition (4) (Section 3.1) whenever $j$ refers to an initial node of the pre-existing conditions:

$$\text{clock}_j = -x,$$

while we assert $\text{clock}_j = 0$ when the index $j$ refers to the new condition.

We evaluated how this can change the recommendation given by our system in a very simple case. First, we ran the simulation as in the preceding section, except that we did not include osteoporosis, and obtained the solution shown in Fig. 5. Then we added back osteoporosis with a starting time 0, while changing the starting time of the remaining four conditions to $-6$. The result is shown in Fig. 6. As it can be seen by comparing the results, the addition of the new condition (osteoporosis) modified the suggested pathway for one of the existing conditions (hypertension). This means that the clinician could consider backtracking and changing the previously established therapy in view of the modified clinical situation. It should be noted that the running time for these examples is substantially the same as that for the example in the previous section ( $\leq 30$ s).

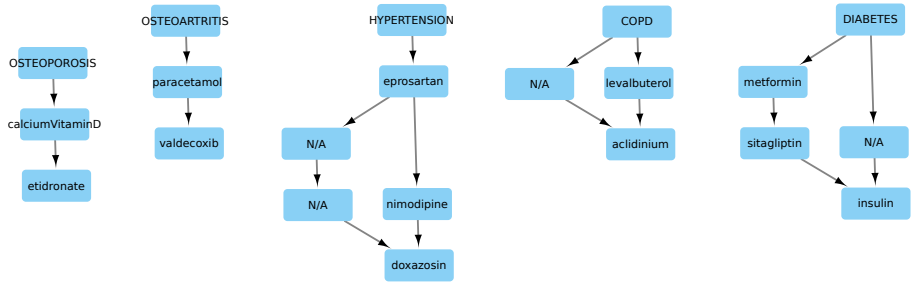

**Fig. 6.** The optimal solution after a fifth condition was added with a time offset.

## 5 Formal Verification

Formulas (2) and (3) correspond to SMT assertions whose number quickly grows even for small graphs. There are more immediate ways of expressing the same problem as an SMT problem, but they turn out to be significantly less efficient. Our idea (extending the approach taken in [6]) is to exploit Isabelle's SMT-LIB generator to automatically produce SMT code from Isabelle definitions that we can formally prove to be correct via formal Isabelle theorems. This SMT code will typically be not as efficient as the SMT code that we will effectively run; however, we can use an SMT solver to prove that the two are equivalent. This allows us to infer (if we trust the SMT solver) that the formal correctness theorems proved in Isabelle apply to the SMT code that we will effectively run.

This approach is illustrated in Fig. 7. While this scheme can be applied generally, we use formulas (2) and (3) (Section 3) to illustrate it. We rewrite them as the following equivalent Isabelle/HOL definitions:

```
abbreviation "conditionTwo' G F ≡
(∀ p. (F p & ¬ isSink' G p) → (∃! c. (G p c & F c)))"

abbreviation "conditionThree' G F ≡
∀ c. (¬ isSource' G c & (∀ p. G p c → ¬ F p)) → (¬ F c)"
```

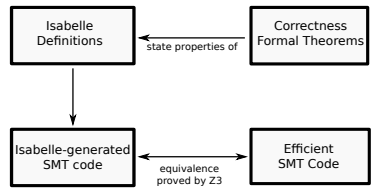

**Fig. 7.** Overview of the formal verification of the SMT code.

Here, $G$ is a graph and $F$ is a set of nodes expressed as predicates (i.e., there is an edge from $m$ to $n$ if and only if $G\ m\ n$ is true, and the node $m$ is selected if and only if $Fm$ is true), and `isSink' G s` is a predicate telling whether $s$ is a sink for the graph $G$; similarly for `isSource'`. To these definitions, we add another, corresponding to the requirement that all the sources must be selected:

```
abbreviation "conditionOne' G F ≡ (∀ s. isSource' G s → F s)"
```

Now, we put together these conditions:

```
abbreviation "formalConditions' G F ≡ conditionOne' G F &
                  conditionTwo' G F & conditionThree' G F"
```

and use Isabelle's SMT generator to produce corresponding SMT code:

```
lemma assumes "formalConditions' G F" shows False
sledgehammer[provers=z3, minimize=false,
             timeout=1, overlord=true] (assms).
```

`sledgehammer` is an Isabelle tool for theorem proving, and it works by negating the thesis of the considered theorem, and to challenge an SMT solver (or other tools) to consider whether the obtained problem is satisfiable. If it is not, it can use the information to find a proof. In this case, we are not interested in theorem proving, but only in the generated SMT code. More details can be found in [6], where we directly use the generated code to compute. Here, we make an indirect use of it: the generated code is used to define a boolean SMT variable `formal`, containing all the corresponding assertions. That is, `formal` will be true if and only if the automatically generated `formalConditions'` G F is true. We do the same for the assertions illustrated at the beginning of Section 3, to obtain an SMT variable that we call `efficient`. In other words, `efficient` will be true if and only if each node$_j$ obeys the group of assertions in Section 3. We add the requirement that $Fj = \text{node}_j$ for any node $j$. If this group assertions and the group of assertions enclosed in `formalConditions'` introduced above were not equivalent, there should be a choice of $F$ satisfying one but not the other group. Therefore, we express the following SMT assertion:

```
(assert (or (and formal (not efficient))
            (and (not formal) efficient))).
```

If the SMT solver returns (`unsat`), this means that the two conditions are indeed equivalent. It should be noted that this check is only needed once; for the example of Section 4, we obtained an (`unsat`) answer in about 5 seconds.

Once we have the guarantee of the equivalence between the SMT code that we run and the SMT code generated from Isabelle definitions, we can start proving formal theorems in Isabelle about these definitions, in order to make sure they present the intended properties.

As an example, we discuss how to prove that the nodes selected by the SMT assertions labelled as `formal` is indeed a path leading from a source to a sink for the corresponding graph. To do so, we start from defining in Isabelle/HOL the canonical notion of walk in a simple directed graph:

```
definition "isWalk G w ≡ w≠[] →
                 (∀i∈{1..<size w}. (w!(i−1),w!i)∈G)",
```

which amounts to asking that any consecutive entries of the node list `w` are joined by an edge (note that `w!0`, `w!1`, ... are the entries of the list `w`). We want to prove

that any selection of nodes of a given graph G obtained from the assertions
formal are the entries of a walk which starts from the source of G, ends in a
sink of G and has no repetitions. To this end, we formally proved the following
theorem:

```
theorem assumes "finite G" "card (sources G)=1"
"irrefl (trancl G)" "conditionOne (set2pred G) (unset F)"
"conditionTwo (set2pred G) (unset F)"
"conditionThree (set2pred G) (unset F)" shows
"∃ w. distinct w & isWalk G w & last w∈sinks G &
              {hd w}=F∩sources G & set w=F∩nodes G"
```

The converter set2pred serves to pass from G represented as a set of pairs to
its representation as a binary boolean predicate. In Isabelle, they have different
types, and using sets is more expressive than using predicates, leading to more
readable statements; on the other hand, sets are not present natively in the
SMT-LIB language, and using predicates is therefore preferable. Similarly, unset
passes from the set F to its representation as a unary boolean predicate.

The theorem assumes that the graph is finite, and that there is only one
source. Given that it is represented as a set of ordered pairs, it is inherently directed, while irrefl (trancl G) states that its transitive closure is not reflexive.
This is the set-theoretical way of requiring that G is acyclic. Under these assumptions, plus the requirements that the SMT solver proved to be equivalent to the
SMT assertions that we execute, the theorem states that the set F∩nodes G is
the entries set of some walk w for G which starts at the source (hd w is the first
entry of w), ends at a sink, and has no repetitions (keyword distinct). The conditions about w in the thesis of the theorem above correspond to requirements
at the point (1) of the list of conditions at the beginning of Section 2.2.

## 6 Related Work

The problem of model composition has been treated by using SMT solvers or constraint solvers in previous works [27, 22, 26, 3, 5]. To the best of our knowledge,
no other existing approach to the problem adopts a metric to be maximised, as
we did here. From an abstract viewpoint, there are some similarities between the
model we proposed in Section 2 and that presented in [13] where, however, there
is only one graph involved, and the semantics associated to it and the problem to
be solved are entirely different. In [11], the specific problem of the composition
of pharmaceutical graphs is solved by the usage of SMT solvers in a way which
inspired the present paper. Here, we addressed two of the main limitations conceded in [11], namely the lack of dosage and timing information, presenting an
SMT-based design which adds those information coordinates to obtain a more
flexible and realistic solution to the problem of minimising conflict in multiple
pathway applications on multimorbid patients.

The issue of verifying SMT code, a solution to which we discussed in Section 5, is underexplored: indeed, SMT solvers are typically used the other way
around, i.e., to aid in the verification of software or in formal proofs [2].

The problem of automated detection and resolution of pathway conflicts in patients with multimorbidity is gaining considerable attention. In [9], ontologies are used to represent pathways, with one additional ontology (Merge Representation Ontology, MRO) created by interviewing clinicians to identify merging criteria, and instantiated to find merge points of two given pathways. No information is given, however, on how adaptable the approach is and whether the instantiation process is automatic. Similarly, recent work proposing extensions to the existing GLARE project [20] also uses ontologies, but focuses on soliciting clinicians' input for conflicts solution, rather than automating it. Michalowski et al. [15] propose constraint logic programming (CLP) to express and deal with conflicts. While CLP solvers and SMT solvers have fundamental similarities, their expressiveness, background technologies and domains of application differ. With the efficiency of the current, mutually competitive, SMT solvers growing at a steadfast pace, their lower expressiveness is getting more and more effectively compensated. Another route has been brought forward in [14], where an ad-hoc algorithm is proposed based on formal rules expressing the actions to be taken upon the happening of given premises expressing conflicts. Whereas this solution has the advantage of accurately solving conflicts, based on the actual medical meaning of the rules themselves, it relies on specific software, which looks problematic in view of extensions and further development.

## 7  Conclusions

We presented a theoretical framework to express the problem of complex model composition as an optimisation problem on a suitably defined score metric, expressed this framework in SMT language and evaluated the whole approach on a concrete, well-known healthcare problem. Additionally, we formally proved the correctness of our SMT code through a novel application of theorem prover Isabelle, exploiting the latter's ability of generating SMT code. This is important because writing SMT code directly is time-consuming and error-prone while, and the existing interfaces of SMT solvers with higher-level languages (e.g., APIs) are currently (as far as we know) not formally verified.

Our approach can be adapted to a wide range of conditions: for example, by imposing manual restrictions on the number of resources, by manually imposing a chosen path in some of the graphs, by changing how $f$ operates, or by offsetting the application of distinct models time-wise; in the latter case, we showed how this can be exploited in a simple way to manage the changes to a patient's therapy following the diagnose of a new condition.

While the current execution time is tolerable for interactive use, we believe that in real situations it can be dramatically lower, on average. Nonetheless, we are experimenting with *scopes*, a feature of Z3 which allows incremental solving by specifying a block of assumptions which are always held, together with a stack of additional assumptions which can be pushed and popped repeatedly. When suitably handled, this feature is likely to allow to reduce the SMT computation time upon modifications of the inputs, since a relevant part of it will remain unchanged throughout the modifications (e.g., the topology of the graphs).

The presented design has some limitations. First, while it can find an execution minimising the conflict, it cannot suggest additional remedies to neutralise or mitigate the arising conflicts. Secondly, it does not handle, currently, directed graph presenting cycles (we are not aware of any existing approach supporting the presence of cycles in our main application domain, e.g., clinical pathways composition). Third, the score calculation could be made context-aware by considering previously occurred resources in the same graphs. A possible attack to the last issue could consist in generalising the maps $f$ and $g$ by adding arguments to them expressing the context: we are conducting experiments in this sense.

# References

[1]  J. Araújo, J. Whittle, and D.K. Kim. "Modeling and composing scenario-based requirements with aspects". In: *Requirements Engineering Conference, 2004. Proceedings. 12th IEEE International*. IEEE Computer Society Press. 2004, pp. 58–67.

[2]  François Bobot et al. "Why3: Shepherd your herd of provers". In: *Boogie 2011: First International Workshop on Intermediate Verification Languages*. 2011, pp. 53–64.

[3]  J. Bowles, B. Bordbar, and M. Alwanain. "A logical approach for behavioural composition of scenario-based models". In: *Formal Methods and Software Engineering: 17th International Conference on Formal Engineering Methods*. Ed. by S. Conchon M. Butler and F. Zaïdi. LNCS 9407. Springer, 2015, pp. 252–269.

[4]  J.K.F. Bowles and B. Bordbar. "A formal model for integrating multiple views". In: *Application of Concurrency to System Design, 2007. ACSD 2007. Seventh International Conference on*. IEEE Computer Society Press, 2007, pp. 71–79.

[5]  J.K.F. Bowles, B. Bordbar, and M. Alwanain. "Weaving true-concurrent aspects using constraint solvers". In: *Application of Concurrency to System Design (ACSD 2016)*. IEEE Computer Society Press, 2016.

[6]  Juliana Küster Filipe Bowles and Marco Bright Caminati. "Mind the gap: addressing behavioural inconsistencies with formal methods". In: *2016 23rd Asia-Pacific Software Engineering Conference (APSEC)*. IEEE Computer Society. 2016.

[7]  Cynthia M Boyd et al. "Clinical practice guidelines and quality of care for older patients with multiple comorbid diseases: implications for pay for performance". In: *Jama* 294.6 (2005), pp. 716–724.

[8]  D. Jackson. *Software Abstractions: logic, language and analysis*. MIT Press, 2006.

[9] Borna Jafarpour and Syed Sibte Raza Abidi. "Merging disease-specific clinical guidelines to handle comorbidities in a clinical decision support setting". In: *Conference on Artificial Intelligence in Medicine in Europe*. Springer. 2013, pp. 28–32.

[10] J. Klein, L. Hélouët, and J.M. Jézéquel. "Semantic-based weaving of scenarios". In: *Proceedings of the 5th international conference on Aspect-oriented software development*. ACM. 2006, pp. 27–38.

[11] A. Kovalov and J. Bowles. "Avoiding medication conflicts for patients with multimorbidities". In: *12th International Conference on Integrated Formal Methods*. LNCS 9681. Springer, 2016, pp. 376–392.

[12] H. Liang et al. "A general approach for scenario integration". In: *MoDELS 2008*. LNCS 5301. Springer, 2008, pp. 204–218.

[13] Michele Lombardi, Michela Milano, and Luca Benini. "Robust scheduling of task graphs under execution time uncertainty". In: *IEEE transactions on computers* 62.1 (2013), pp. 98–111.

[14] Joan Albert López-Vallverdú, David Riaño, and Antoni Collado. "Rule-based combination of comorbid treatments for chronic diseases applied to hypertension, diabetes mellitus and heart failure". In: *Process Support and Knowledge Representation in Health Care*. Springer, 2013, pp. 30–41.

[15] Martin Michalowski et al. "Using constraint logic programming to implement iterative actions and numerical measures during mitigation of concurrently applied clinical practice guidelines". In: *Conference on Artificial Intelligence in Medicine in Europe*. Springer. 2013, pp. 17–22.

[16] L. De Moura and N. Bjørner. "Z3: An efficient SMT solver". In: *TACAS 2008*. Vol. 4963. LNCS 4963. Springer, 2008, pp. 337–340.

[17] Tobias Nipkow, L. C. Paulson, and Markus Wenzel. *Isabelle/HOL: a proof assistant for higher-order logic*. London, UK: Springer-Verlag, 2002. ISBN: 3-540-43376-7.

[18] OMG. *Business Process Model and Notation. Version 2.0*. Document id: formal/2011-01-03. OMG. http://www.omg.org., 2011. URL: http://www.omg.org.

[19] OMG. *UML: Superstructure. Version 2.4.1*. Document id: formal/2011-08-06. OMG. http://www.omg.org., 2011. URL: http://www.omg.org.

[20] Luca Piovesan, Gianpaolo Molino, and Paolo Terenziani. "An ontological knowledge and multiple abstraction level decision support system in healthcare". In: *Decision Analytics* 1.1 (2014), p. 1.

[21] R.Reddy et al. "Composing sequence models using tags". In: *Proc. of MoDELS Workshop on Aspect Oriented Modeling*. 2006.

[22] J. Rubin, M. Chechik, and S.M. Easterbrook. "Declarative approach for model composition". In: *MiSE 2008*. ACM, 2008, pp. 7–14.

[23] Paul Shannon et al. "Cytoscape: a software environment for integrated models of biomolecular interaction networks". In: *Genome research* 13.11 (2003), pp. 2498–2504.

[24] S. Uchitel, G. Brunet, and M. Chechik. "Synthesis of partial behavior models from properties and scenarios". In: *IEEE Transactions on Software Engineering* 35.3 (2009), pp. 384–406.
[25] J. Whittle, J. Araújo, and A. Moreira. "Composing aspect models with graph transformations". In: *Proceedings of the 2006 international workshop on Early aspects at ICSE*. ACM. 2006, pp. 59–65.
[26] M. Widl et al. "Guided merging of sequence diagrams". In: *SLE 2012*. LNCS 7745. Springer, 2013, pp. 164–183.
[27] D. Zhang, S. Li, and X. Liu. "An Approach for Model Composition and Verification". In: *NCM 2009*. IEEE Computer Society Press., 2009, pp. 1102–1107.